\documentclass[twocolumn,floatfix]{aastex63}

\usepackage{amssymb}
\usepackage{amsmath}
\usepackage{hyperref}

\usepackage{graphicx}

\begin{document}

\title{The Earliest Stage of Galactic Star Formation}

\correspondingauthor{Charles Steinhardt}
\email{steinhardt@nbi.ku.dk}

\author[0000-0003-3780-6801]{Charles L. Steinhardt}
\affiliation{Cosmic Dawn Center (DAWN)}
\affiliation{Niels Bohr Institute, University of Copenhagen, Lyngbyvej 2, DK-2100 Copenhagen \O}

\author[0000-0001-7633-3985]{Vadim Rusakov}
\affiliation{Cosmic Dawn Center (DAWN)}
\affiliation{Niels Bohr Institute, University of Copenhagen, Lyngbyvej 2, DK-2100 Copenhagen \O}

\author[0000-0003-3873-968X]{Thomas H. Clark}
\affiliation{California Institute of Technology, 1200 E. California Blvd., Pasadena, CA 91125, USA}
\affiliation{Cosmic Dawn Center (DAWN)}

\author[0000-0002-6459-8772]{Andrei Diaconu}
\affiliation{California Institute of Technology, 1200 E. California Blvd., Pasadena, CA 91125, USA}
\affiliation{Cosmic Dawn Center (DAWN)}

\author[0000-0002-1975-4449]{John Forbes}
\affiliation{Flatiron Center for Computational Astrophysics, 162 5th Ave 9th floor, New York, NY 10010, USA}

\author[0000-0003-0639-025X]{Conor McPartland}
\affiliation{Cosmic Dawn Center (DAWN)}
\affiliation{Niels Bohr Institute, University of Copenhagen, Lyngbyvej 2, DK-2100 Copenhagen \O}

\author[0000-0002-5460-6126]{Albert Sneppen}
\affiliation{Cosmic Dawn Center (DAWN)}
\affiliation{Niels Bohr Institute, University of Copenhagen, Lyngbyvej 2, DK-2100 Copenhagen \O}

\author[0000-0003-1614-196X]{John Weaver}
\affiliation{Cosmic Dawn Center (DAWN)}
\affiliation{Niels Bohr Institute, University of Copenhagen, Lyngbyvej 2, DK-2100 Copenhagen \O}


\begin{abstract}
Using a recently-developed technique to estimate gas temperatures ($T_\textrm{SF}$) in star-forming regions from large photometric surveys, we propose a diagram, analogous to the Hertzsprung-Russell diagram for individual stars, to probe the evolution of individual galaxies.  On this $T_\textrm{SF}$-sSFR (specific star formation rate) diagram, a small fraction of star-forming galaxies appear to be dominated by different feedback mechanisms than typical star-forming galaxies. These galaxies generically have younger stellar populations, lower stellar masses and increase in relative abundance towards higher redshifts, so we argue that these objects are in an earlier stage of galactic star formation. Further, Hubble observations find that these ``core-forming'' galaxies also exhibit distinct morphology, and that tracks on the $T_\textrm{SF}$-sSFR diagram are also a morphological sequence. Thus, unlike starburst phases which can be triggered environmentally, these earliest, core-forming galaxies, appear to be a stage that typical galaxies go through early in their star formation history. We therefore argue that most galaxies first go through a core formation stage, then subsequently disk formation, and finally become quiescent. 
\end{abstract}


\section{Introduction}
\label{sec:intro}

It is perhaps natural to divide the history of a typical galaxy into three phases, each dominated by different astrophysical processes.  During the first phase, gravity brings the ingredients which will become a protogalaxy into close proximity.  Then, a wide variety of baryonic processes assemble those ingredients into stars, a central supermassive black hole, and other components of a galaxy.  Finally, for most high-mass galaxies, at some point in the past these formation processes have predominantly stopped due to processes which have not yet been conclusively determined, leaving a quiescent galaxy with a passive galactic nucleus.  

However, at present only the latter two of these classes have been observed.  A variety of methods \citep{Williams2009,Muzzin2013,Steinhardt2020} are capable of unambiguously labeling all but a small fraction of galaxies as either star-forming or quiescent in large photometric catalogs.  The small fraction of galaxies with uncertain classifications often lie in the ``green valley'' \citep{Martin2005,Salim2014}, considered a possible transition stage between bluer star-forming galaxies and redder quiescent ones.

Most star-forming galaxies exhibit remarkably similar growth, including tight relationships at fixed redshift between stellar mass and SFR (the star-forming main sequence; \citealt{Noeske2007,Peng2010,Speagle2014,Schreiber2018}), mass and metallicity \citep{Tremonti2004,Sanders2020} and mass and radius \citep{vanderWel2014}.  If these relations are the end result of various feedback mechanisms \citep{Peng2010}, galaxies in a transition phase between initial assembly and subsequent main sequence-like evolution might not follow them.  This anomalous behavior might even provide a means of identifying these galaxies.

Additional growth phases that do not follow these relations (or lie at the extremes) have been proposed or observed.  Galaxies with enhanced star formation rates (SFR) are often described as starbursts, and lower-redshift populations of ultra-luminous infrared galaxies (ULIRGs; \citealt{Lonsdale2006}) may be related to starbursts.  Simulations also suggest that there may be two star-forming phases, one burst-like and the other steady \citep{Stern2021}.  Although every galaxy is expected to have a steady star-forming phase and a quiescent phase, bursts are believed to be associated with mergers or other environmental factors \citep{RodriguezMontero2019}, and thus not every galaxy will have one.  Therefore, starbursts are not considered strong candidates as the earliest evolutionary phase of typical galaxies.

In this work, the IMF temperature ($T_{\textrm{SF}}$) - specific star formation rate (sSFR) diagram, a galaxy analogue of the Hertzsprung-Russell diagram \citep{Hertzsprung1909,Russell1914} for stars, is introduced and used to produce galactic evolutionary tracks and identify galaxies in the earliest stages of star formation.  These earliest galaxies exhibit distinctive properties and morphology, with a gradual transition towards more typical main sequence behavior indicating that this is an evolutionary sequence.   In contrast to proposed burst phases, this new, core-forming phase is seemingly an early, steady star-forming state shared by all galaxies and transitions gradually into the more familiar main sequence. Furthermore, if most galaxies follow a similar evolutionary track, there will also be local and low-redshift analogues of this pre-main sequence phase.  

In \S~\ref{sec:method}, the methodology behind measuring $T_{\textrm{SF}}$ from photometric catalogs is summarized, as well as several tests performed to validate these temperatures.  The resulting $T_{\textrm{SF}}$-sSFR diagram is developed in \S~\ref{sec:sSFRTplane}, along with the conclusion that the earliest stage of galaxy evolution should consist of a core-forming phase.  Finally, these results are discussed in \S~\ref{sec:discussion}, including additional predictions and a toy model providing a possible astrophysical interpretation.

\section{Measuring Gas Temperatures from Photometric Surveys}
\label{sec:method}

This work relies on a recently-developed technique for inferring the temperature of gas in star-forming molecular clouds from multi-wavelength photometric surveys.  It is expected that the IMF should depend upon the temperature of gas in star-forming molecular clouds \citep{LyndenBell1976,Larson1985,Bernardi2017,Jermyn2018}.  Thus, a measurement of the IMF might also provide a measurement of gas temperatures during the last major epoch of star formation.  

The IMF is expected to depend not only on gas temperature, but on a complex series of interactions sensitive to metallicity, density, pressure, and environment \cite{Larson1985}.  As a result, the Galactic IMF is determined observationally rather than theoretically, despite the inherent difficulty of these observations resulting in several common approximations \citep{Salpeter1955,Kroupa2001,Chabrier2003}.  The Kroupa IMF \citep{Kroupa2001} has two breakpoints, one determined by a Jeans mass-like calculation relating to the initial collapse of a cloud and the second by a similar calculation during fragmentation as the cloud collapses.  With all other conditions held fixed, it is then possible to derive a family of IMFs which depends on the gas temperature in star-forming regions, $T_{SF}$ \citep{Jermyn2018}:

\begin{equation}
    \frac{dN}{dm} \propto 
    \begin{cases}
       m^{-0.3} & m < 0.08 M_{\odot} f(T_{SF}) \\
       m^{-1.3} & 0.08 M_{\odot} f(T_{SF}) < m  < 0.5 M_{\odot} f(T_{SF}) \\
       m^{-2.3} & 0.5 M_{\odot} f(T_{SF}) < m, \\
    \end{cases}
    \label{eq:kroupa_2}
\end{equation} 

Although there is broad agreement that higher temperature lead to higher-mass breakpoints, the exact temperature dependence is more difficult to establish.  Various studies have concluded that the break masses could scale as $f(T_{SF}) \propto T_{SF}$ \citep{Hopkins2012}, $f(T_{SF}) \propto T_{SF}^{3/2}$\citep{Jeans1902}, $f(T_{SF}) \propto T_{SF}^2$ \citep{Jermyn2018}, or $f(T_{SF}) \propto T_{SF}^{5/2}$ \citep{Chabrier2014}.  Here, we select the $T_{SF}^2$ dependence with the normalization set to match a Kroupa IMF at $T_{SF} = 20$K (which is the characteristic temperature of Galactic star-forming clouds) following the convention established in \cite{Jermyn2018} and \cite{Sneppen2022}

IMFs corresponding to a grid of gas temperatures, ranging from 10-60K, are then used to generate synthetic spectra using the Flexible Stellar Population Synthesis (FSPS) libraries \citep{fsps} for use with the EAZY photometric template fitting code \citep{Brammer2008}.  EAZY uses a basis of 12 templates generated as linear combinations of 560 individual models with various ages, star formation histories, metallicities, and extinction.  A set of 12 basis templates is generated for each $T_{SF}$.  The best-fit reconstructed spectrum is then found for every object in the COSMOS2015 photometric catalog using each IMF, and the lowest $\chi^2$ across all IMFs is selected as the best fit.  

Although in principle this allows an IMF determination for all $\sim 10^6$ COSMOS2015 galaxies, in practice this is not possible because of the strong covariances between the IMF, metallicity, and extinction \citep{Sneppen2022}.  Since they are not fully degenerate, given sufficient information it is possible to distinguish between, e.g., a bluer spectrum due to a top-heavier (or bottom-lighter) IMF and due to lower extinction.  However, this requires high signal-to-noise measurements across multiple bands.  In order to explore this, a mock dataset was constructed at a range of IMFs and observed in the COSMOS2015 filters.  It was found that the gas temperature could be accurately recovered for the $\sim 10\%$ brightest objects in COSMOS2015, but that the remainder would require additional observations to determine the IMF \citep{Sneppen2022}.  The work here relies only on those $\sim 14\times10^4$ objects, which is the cause for low statistics at $z \gtrsim 2$. 

In addition to these strong covariances, there is a complete mathematical degeneracy between the IMF and the star formation history (SFH), as an identical stellar population can be produced with different combinations of IMF and SFH.  Mock datasets were used to show that the template fitting technique used here is primarily sensitive to the highest-mass break in the {\em stellar population}.  For a continuous SFH, as is likely for typical star-forming galaxies, this should correspond to the highest-mass break in the IMF, and therefore can be used as an IMF and gas temperature indicator.  However, in the case of rapid quenching, this break can instead come from the SFH, and a population of such objects is discussed in \citet{Steinhardt2022b} in connection with star formation turnoff.  These uncommon objects often exhibit two local minima in the minimum $\chi^2(T_{SF})$, with one minimum corresponding to the gas temperature (and lying on the main locus in the $T_{SF}-$sSFR diagram) and the other corresponding to the break in SFH.  

Finally, it is necessary to consider other effects on the IMF apart from gas temperature.  A Jeans-mass like calculation is sensitive to anything which changes either gravitational or thermodynamical effects.  In particular, it is expected that the IMF should be sensitive to metallicity in addition to temperature, and it is known that galactic metallicities change with redshift.  Given the difficulty of constraining template fits with addition of a single-parameter family of IMFs, is it not possible to fit a multiple-parameter family of IMFs with current photometric surveys.  Likely, the correct interpretation of $T_{SF}$ is instead as a combination of gas temperature, metallicity, and other parameters.  

\begin{figure*}[!ht]
\begin{center}
\includegraphics[width=1\linewidth]{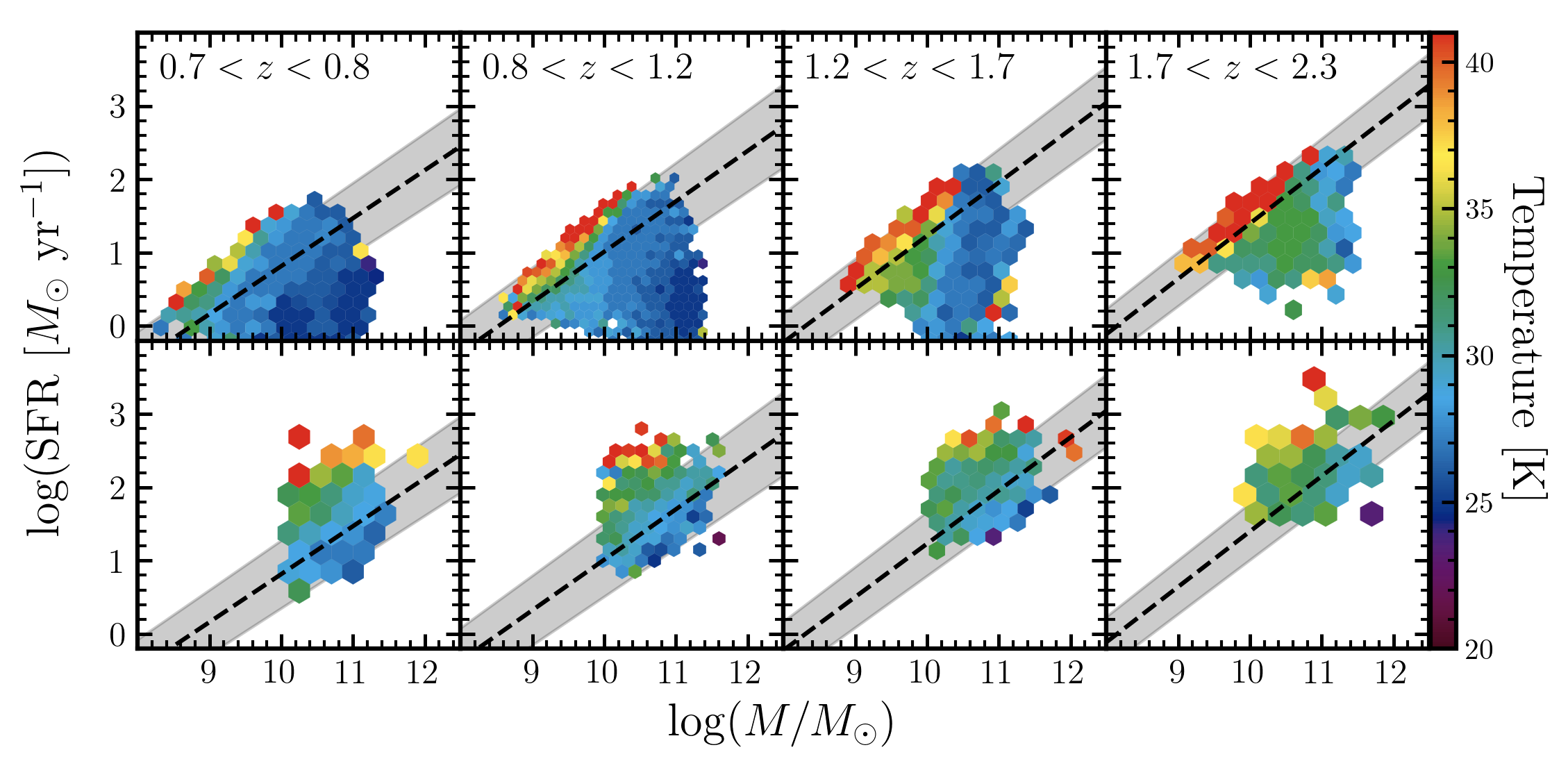}
\caption{Gas ($T_{SF}$; top) and dust (bottom) temperatures for COSMOS galaxies as a function of stellar mass $M_{\star}$ and SFR in four redshift windows.  The $T_{SF}$ are adapted from \citet{Steinhardt2022a}.  The dust temperature measurements are based on the analysis of the far-infrared and sub-millimeter stacked observations made with the {\it Herschel Space Observatory} and are reproduced (B. Magnelli, private communication) from \citet{Magnelli2014}.  For ease of comparison, all panels use the Galactic IMF-derived COSMOS2015 stellar masses and SFR, even though a modified IMF will also modify both star formation rates and stellar mass.  The dashed lines show the best-fit main-sequence ridge lines with the shaded region showing one standard deviation uncertainty from \cite{Speagle2014}.  Both gas and dust temperatures exhibit a similar gradient, increasing towards lower stellar masses and higher SFR at all redshifts.  The two measurements are further in approximate quantitative agreement, despite the two methods each having significant but independent sources of uncertainty.  This is consistent with an equilibrium between cool molecular gas and dust temperatures, as well as the interpretation that $T_{SF}$ is indeed a proxy for temperature.  }
\label{fig:tempcomp}
\end{center}
\end{figure*}

Nevertheless, there are several indications that interpreting $T_{SF}$ as gas temperature is reasonable \citep{Sneppen2022}.  Notably, a comparison between $T_{SF}$ and dust temperatures shows strong agreement (Fig.~\ref{fig:tempcomp}) in not only the broad qualitative trends but even quantitatively \citep{Steinhardt2022a}.  Thus, $T_{SF}$ proves a reasonable proxy for (cool) dust temperature, and if gas and dust are approximately in equilibrium, $T_{SF}$ should then indicate gas temperature in star-forming regions.  The resulting change in stellar masses at $z \sim 2$ produces a picture more consistent with downsizing and quenching at lower redshifts \citep{Steinhardt2022b}.  This work, showing that an evolutionary sequence selected entirely from template fitting is also a morphological sequence, further validates that $T_{SF}$ is a meaningful astrophysical indicator.

\section{The $T_{\textrm{SF}}$-sSFR diagram}
\label{sec:sSFRTplane}

The key diagnostic developed in this work is the $T_{\textrm{SF}}$-sSFR diagram.  The specific star formation rate, or star formation rate per unit mass, is a measure of how efficiently a galaxy is turning material into new stars.  The IMF temperature $T_{\textrm{SF}}$ \citep{Sneppen2022,Steinhardt2022c,Steinhardt2022b,Steinhardt2022a}, is an estimate of the temperature in star-forming molecular clouds at the time of star formation.  Thus, a diagram comparing $T_{\textrm{SF}}$ and sSFR is intended as an analogue of the Hertzsprung-Russell (H-R) diagram comparing $T$ and luminosity for individual stars.

$T_{\textrm{SF}}$ is also a measure of the Jeans mass \citep{Jeans1902} necessary for these clouds to collapse and form stars, so it should be expected that there is a relationship between sSFR and $T_{\textrm{SF}}$.  Using the recently-developed techniques described in \citet{Sneppen2022}, it is possible to determine this relationship for the first time, then use it as a probe of the astrophysics in star-forming galaxies.

It should be stressed that both sSFR and $T_{\textrm{SF}}$ are inferred quantities and rely on a large set of assumptions.  It is well established that star formation rates (and therefore sSFR as well) are difficult to constrain in photometric template fitting \citep{Brammer2008,Conroy2009b,Ilbert2013,Speagle2014,Laigle2016,Weaver2022}.  The addition of a new parameter to measure the shape of the IMF and thus infer $T_{\textrm{SF}}$ adds additional potential degeneracies \citep{Sneppen2022}.  Further, the remainder of this work asserts that the shape of the parametrized Kroupa IMF is a proxy for $T_{\textrm{SF}}$, as based on the Jeans mass approximation, using a specific scaling out of several options, as described in \S~\ref{sec:method}.

Finally, this method assumes that galaxies can be described with a single, best-fit $T_{\textrm{SF}}$, even though it is likely that molecular clouds in star-forming regions will have a range of gas temperatures. As a result, even if there is truly an underlying relationship between star formation efficiency and gas temperature, the best-fit sSFR and $T_{\textrm{SF}}$ for a galaxy containing molecular clouds in a wide range of conditions will not necessarily lie on that relation.  However, as it is commonly done in constructing photometric catalogs and template fitting, it is necessary to approximate galaxies as being effectively monolithic.  
They are assumed to be well-characterized by a single gas temperature for most star-forming clouds and a single sSFR which describes the efficiency in all of those regions.  

\subsection{Identifying Three Primary Growth Regimes}
\begin{figure*}[!ht]
\begin{center}
\includegraphics[width=1\linewidth]{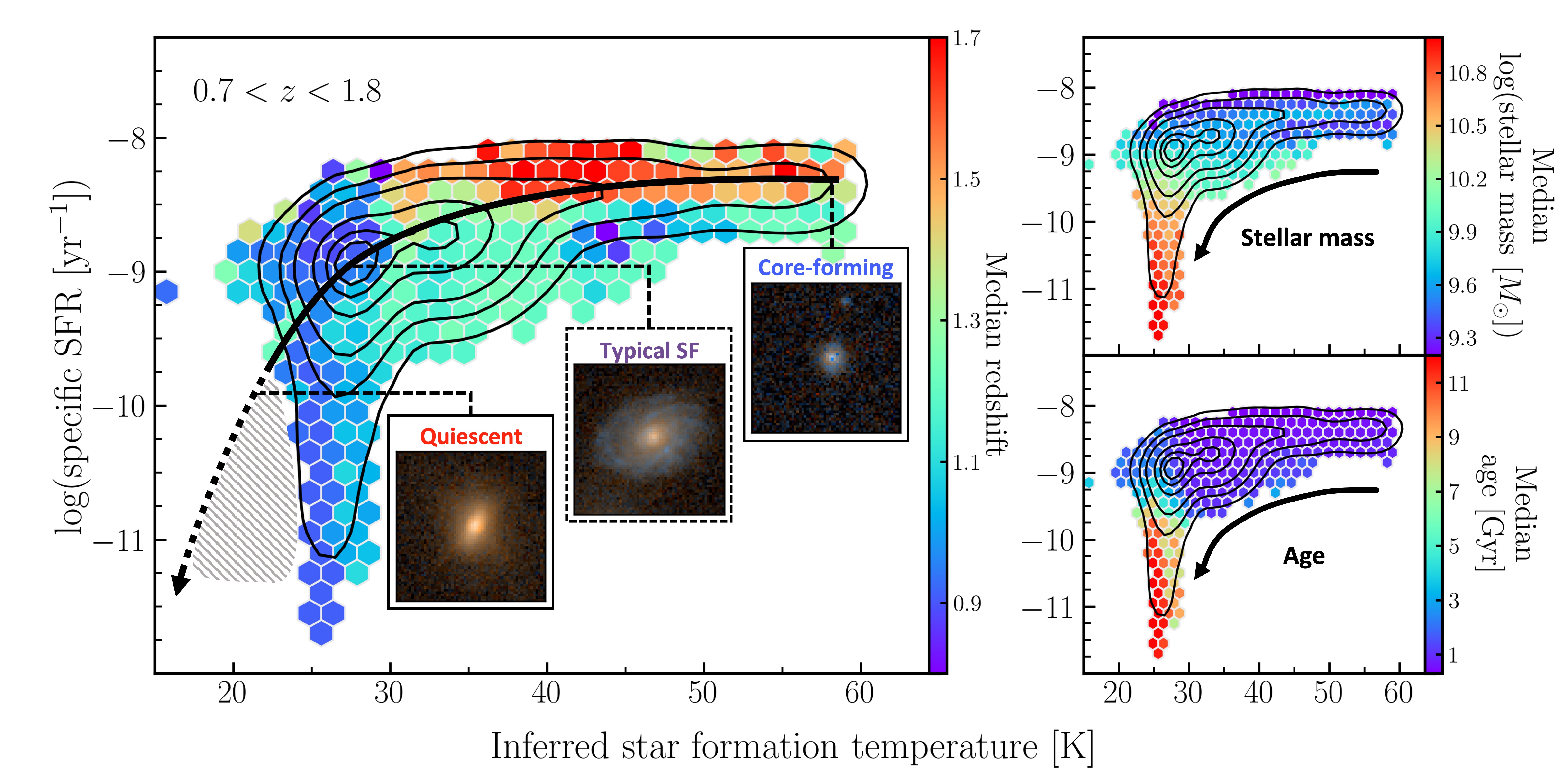}
\caption{The $T_{\textrm{SF}}$-sSFR diagram, analogous to a Hertzsprung-Russell diagram for galaxies, for COSMOS2015 galaxies at $0.7 < z < 1.8$ showing bins with at least 10 data points.  Panels are colored by the median (left) redshift; (top-right) stellar mass; and (bottom-right) age of the stellar population.  Moving from top-right to bottom left, galaxies become more massive, with older stellar populations, and are more commonly observed at later times.  Although no morphological information was used to produce the diagram, this is also a morphological sequence (see also Figure \ref{fig:cutouts}).  The diagram is therefore interpreted as an evolutionary track followed by typical galaxies, with core-forming galaxies identified as the initial stage of star formation.  The dashed region in the left panel indicates that the IMF cannot be recovered at $T_{\textrm{SF}} \lesssim 25$ K \citep{Sneppen2022}, although observational evidence (see \S~\ref{sec:IMFvariations}) indicates that the IMF may be even bottom-heavier than $T_{\textrm{SF}} \sim 25$ K.} 
\label{fig:diagram}
\end{center}
\end{figure*} 
The distribution of galaxies on the $T_{\textrm{SF}}$-sSFR diagram (Fig. \ref{fig:diagram}) exhibits a continuous locus which can be described as three different behaviors, labeled as core-forming, typical star-forming, and quiescent.  At fixed redshift, these groups also comprise a mass sequence, with the lowest-mass galaxies selected as core-forming and the highest-mass as quiescent.  Although two most distinctive relationships are in the high-$T_{\textrm{SF}}$ (core-forming) and low-sSFR (quiescent) tails of the distribution, the vast majority of objects in the COSMOS catalog, as the name suggests, are selected as typical star-forming.  

The three groups are characterized by their different relationships between sSFR and $T_{\textrm{SF}}$ (Fig. \ref{fig:diagram}):
\begin{enumerate}
    \item {\textit{Core-forming galaxies} have a wide range of $T_{\textrm{SF}}$, with an increase in $T_{\textrm{SF}}$ corresponding to only a small increase in sSFR.}
    \item {\textit{Normal star-forming galaxies}, the bulk of the population, have a relatively narrow range of both sSFR and $T_{\textrm{SF}}$.  The range is perhaps to narrow to effectively constrain the relationship, but the increase in sSFR with temperature is far sharper than for core-forming galaxies.}
    \item {\textit{Quiescent galaxies} have a wide range of sSFR, always lower than for star-forming galaxies, but are all nearly at identical $T_{\textrm{SF}}$.}
\end{enumerate}

\subsection{Variations in the IMF Over Galactic Lifetimes} \label{sec:IMFvariations}

The tracks indicated on the $T_{\textrm{SF}}$-sSFR diagram correspond to a range of different IMFs over the lifetime of a galaxy.  The earliest, core-forming stages would be characterized by the bottom-lightest IMFs, with the IMF becoming progressively bottom-heavier over time.  The eventual stellar population would be combination of all of these IMFs, although the most recent star formation (and thus the bottom-heaviest IMF) would dominate the observed light.  

This mixture of different stellar populations might provide a possible explanation for the seemingly-conflicting combination of current observational measurements constraining galactic IMFs.  Every technique has both large statistical uncertainties and potentially larger systematic uncertainties, and thus is individually inconclusive.  Nevertheless, studies focusing on nearly every individual region of the stellar mass function appear to report an excess when compared with expectations taken from a Galactic IMF.  Black hole merger rates from LIGO find an excess compared to predictions using a Galactic IMF, consistent with a top-heavier IMF that produced more black holes \citep{Kovetz2018,Flitter2021}.  Delay-time distributions for Type Ia supernovae indicate a possible excess of white dwarfs, which would indicate an intermediate-heavier IMF.  Finally, spectral lines indicate an excess of  low-mass stars in early-type galaxies compared with a Galactic IMF \citep{Conroy2012}, a similar result to studies of the Gaia-Enceladus population believed to have been a dwarf galaxy prior to merging with the Milky Way \citep{Hallakoun2021}.

Each of these results, in addition to probing a different stellar mass range, also probes a different time period.  Thus, an evolutionary track that moves from an initial, bottom-light IMF to a progressively bottom-heavier IMF over time is potentially consistent with all of these results.  The black hole-black hole merger excess in LIGO lies towards higher redshift.  Similarly, the time delay measurements indicate an excess of white dwarfs formed at early times.  However, early-type galaxies are observed very late in their evolutionary history, towards the bottom-left corner of the $T_{\textrm{SF}}$-sSFR diagram.

In order to be consistent with the \citet{Conroy2012} and \citet{Hallakoun2021} results, quenching galaxies must not merely have a Galactic-like IMF; but must actually become bottom-heavy in the last stages of star formation.  The gas temperature measurements presented here find a best-fit $T_{\textrm{SF}}$ of $\sim 25$K, corresponding to an IMF that is instead slightly bottom-lighter than the Milky Way.  However, we note that a comparison with mock datasets finds that this technique is primarily sensitive to bottom-light populations (\citealt{Sneppen2022}, Fig. 13), since those produce the most distinctive shapes in rest-frame UV photometry.  As the blue stars disappear, whether via aging of the stellar population or via a bottom-heavier IMF, the covariances become too strong to constrain the IMF, and the best-fit IMF by default ends up being approximately Galactic.  Thus, a measurement of $T_{\textrm{SF}} = 25$K is more properly a measurement that $T_{\textrm{SF}} \lesssim 25$K.  If galaxies in the bottom-left corner of the $T_{\textrm{SF}}$-sSFR diagram instead have $T_{\textrm{SF}} \sim 10-15$K, which is still above the contribution from massive stars (cf. \citealt{Papadopoulos2010}), such an evolutionary track could produce all of the seemingly-conflicting observational results.
\section{Discussion}
\label{sec:discussion}

With the addition of a parameter corresponding to changes in the IMF due to the gas temperature in star-forming regions, it is possible to measure that gas temperature, $T_{\textrm{SF}}$, in large photometric catalogs.  An examination of the relationship between specific star-formation rate and the best-fit temperature finds three distinct behaviors.  Core-forming galaxies exhibit a wide range of $T_{\textrm{SF}}$ at relatively similar sSFR.  Typical star-forming galaxies exhibit a narrow range of both parameters at fixed redshift, consistent with previous studies \citep{Noeske2007,Peng2010,Speagle2014,Schreiber2018,Tremonti2004,Sanders2020}.  Finally, quiescent galaxies exhibit nearly constant $T_{\textrm{SF}}$ at a wide range of low sSFR.  Previous studies have typically separated galaxies into star-forming and quiescent, but the core-forming branch is a new class of objects that arise from measuring $T_{\textrm{SF}}$.

\subsection{Core-forming Galaxies and ``Blue Nuggets''}

The core-forming galaxies bear strong resemblance to a population dubbed blue nuggets\footnote{It is likely that some of the blue nuggets have been identified as ``green pea'' galaxies at $0.1 < z < 0.4$ \citep{cardamone2009}.  Their physical properties appear to be the same, but their identification as distinct galaxies is based on their strong nebular emission observed in the SDSS r-band.} most frequently found at $1 < z < 2$ \citep{huertas-company2018,lapiner2023}.  They have similarly high sSFR, which places them at the top of the star-forming main sequence, small stellar masses, young stars, little dust obscuration and compact morphologies, as seen in data \citep{cardamone2009,huertas-company2018} and simulations \citep{zolotov2015,Tacchella2016b,tacchella2016a,lapiner2023}.  The increased abundance of core-forming galaxies towards $1 < z < 2$ also supports this comparison.  However, best-fit S\'{e}rsic indices of the core-forming galaxies indicate that they are predominantly smaller versions of the typical star-forming galaxies, consistent with a bulge+disk structure with the index of $n \sim 1$ or irregular shape due to mergers (Fig.~\ref{fig:sersic}), rather than singular compact bulges.  In fact, the $R_{20}$ radii show that the core-forming galaxies are just as diffuse in the center as the typical main-sequence galaxies are (with $n \sim 1$), in contract to very compact bulges in quiescent galaxies.  It is perhaps worth noting that only one of $R_{20}$, $R_{50}$, and $R_{80}$ would be insufficient to see the entire picture, a result which appears to apply more broadly to galaxy morphological studies (cf. \citealt{Mowla2019}).

As the morphological quantities suggest, most core-forming galaxies appear to be distinct from the population of blue nuggets.  It is however possible that these galaxies are in a post-blue-nugget phase characterized by formation of a blue disk once the bulge is formed.  In either case, the photometric procedure used here is effective for differentiating between the typical disk galaxies and their structurally similar smaller versions, which would otherwise be indistinguishable.

\subsection{Interpretation as an Evolutionary Sequence}

These three groups form a mass sequence at fixed redshift and display a progression in the best-fit age of their stellar populations (Fig. \ref{fig:diagram}). The gradients in both parameters across galaxy populations indicate an evolutionary sequence between core-forming, normal star-forming and quiescent galaxies. In addition, there is a redshift dependence to their relative number densities: at higher redshift, there are more core-forming and fewer quiescent galaxies. This is consistent with numerous other studies finding mass downsizing (cf. \citealt{Fontanot2009}) in galaxy evolution. It is therefore natural to consider core-forming galaxies as some sort of early star-forming phase, prior to typical star formation which is far more prevalent in photometric surveys.

\subsection{Interpretation as Various Feedback Modes}
\label{sec:feedback}

A model of the various feedback mechanisms responsible for the star-forming main sequence is well beyond the scope of this work.  Here, a toy model is proposed which might explain the two tails (core-forming and quiescent) of the distribution. \newline  

\textbf{Typical Star-Forming:} Several models of the main sequence invoke tight feedback between conditions in star-forming molecular clouds and high-mass (and therefore newly-formed) stars.  For example, cosmic rays from the deaths of massive stars might provide the dominant heating mechanism responsible for setting gas temperatures \cite{Papadopoulos2010}, but an increase in SFR and thus higher gas temperatures would also increase the Jeans mass and eventually regulate the SFR.  At fixed gas density this would produce an equilibrium solution, and as gas density declines, an attractor solution exists \citep{Steinhardt2020b}.  

A variety of different mechanisms have been proposed, including gas regulation \citep{Peng2010,Lilly2013,Peng2015}, cosmic ray feedback \citep{Papadopoulos2010,Steinhardt2020b}, and even stochastic processes as an alternative to feedback \citep{Kelson2014}.  However, it has not yet been possible to observationally determine which, if any, of these mechanisms is correct.  A possible entry point would be the discovery of galaxies transitioning from their initial assembly to main sequence-like growth, before these feedback processes have had time to dominate their observed properties. 

In this picture, the two tails of the distribution would then be produced by conditions which break one of these two feedback modes.  \newline 

\textbf{Quiescent:} As sSFR declines, at some point young stars will no longer provide the dominant contribution to gas temperature.  For example, given the current Galactic SFR of $\sim 1 M_\odot/\textrm{yr}$, molecular cloud temperatures are insensitive to small changes in SFR.  Thus, the quiescent tail of the distribution would asymptotically approach a constant $T_{\textrm{SF}}$ as sSFR goes to zero. Here, $T_{\textrm{SF}}$ decouples as it is a measure of the gas temperature at the time when the stellar population is formed and not the current temperature \citep{Sneppen2022,Steinhardt2022b}. 

A constant $T_{\textrm{SF}}$ thus implies that by the time galaxies start to quench, contributions from star formation have already ceased to dominate gas temperatures.  This would be consistent with proposed mechanisms such as strangulation \citep{Peng2015} or gas depletion \citep{Cortese2021}.  However, a different behavior would be expected if quenching were driven by major mergers or some other external event which could occur even at high sSFR. \newline

\textbf{Core-Forming:} The core-forming galaxies can perhaps be explained by breaking the other feedback mode.  If a galaxy has sufficiently high gas density, an increase in temperature and the corresponding Jeans mass will have negligible effect on SFR.  Rather, the limiting factor might be infall rates or other effects on gas availability.  

Young, massive stars would still provide the dominant source of gas heating, but the relevant equilibrium would be set by cooling mechanisms rather than by feedback limiting SFR.  For example, if the predominant cooling mechanism is black-body radiation, $\textrm{sSFR} \propto T_{SF}^4$, or equivalently, $T_{SF} \propto \textrm{sSFR}^{1/4}.$ 

\subsection{Inside-Out Growth}

This explanation for early main sequence galaxies additionally predicts that their star-forming regions must be compact.  Due to the high temperatures, stars can only form in high-density regions where the increased Jeans mass does not inhibit collapse.  If baryons initially are densest in the central regions of galaxies, then stars will only form in those central regions.  As $T_{SF}$ decreases, lower density regions further out will be able to form stars.  

Perhaps this suggests that the two star-forming phases identified here correspond to distinct regions.  In their initial stages, star-forming galaxies can only form stars in central regions, even if the morphological features of a bulge may not yet be present.  Thus, we observe protogalaxies to be remarkably compact \citep{vanderWel2014}, and observe the oldest Galactic stellar populations towards the Galactic center \citep{Baade1944,Trippe2008}.  It is for this reason that the high-$T_{\textrm{SF}}$ branch is labeled in this work as ‘core-forming'.

Once the central region has finished forming stars, no high-density regions remain and star formation rates decline, decreasing $T_{\textrm{SF}}$.  Normal star-forming galaxies, corresponding to most of the star-forming main sequence, then form stars in the lower-density disk.  This is consistent with hints that the star-forming main sequence, when viewed in terms of disk mass rather than the full stellar mass, may be redshift-independent \citep{Abramson2014}.  Finally, quiescent galaxies have even ceased to efficiently form disk stars.

\subsection{Interpretation as a Morphological Sequence}
\label{sec:morphology}

\begin{figure*}[!ht]
\begin{center}
\includegraphics[width=1\linewidth]{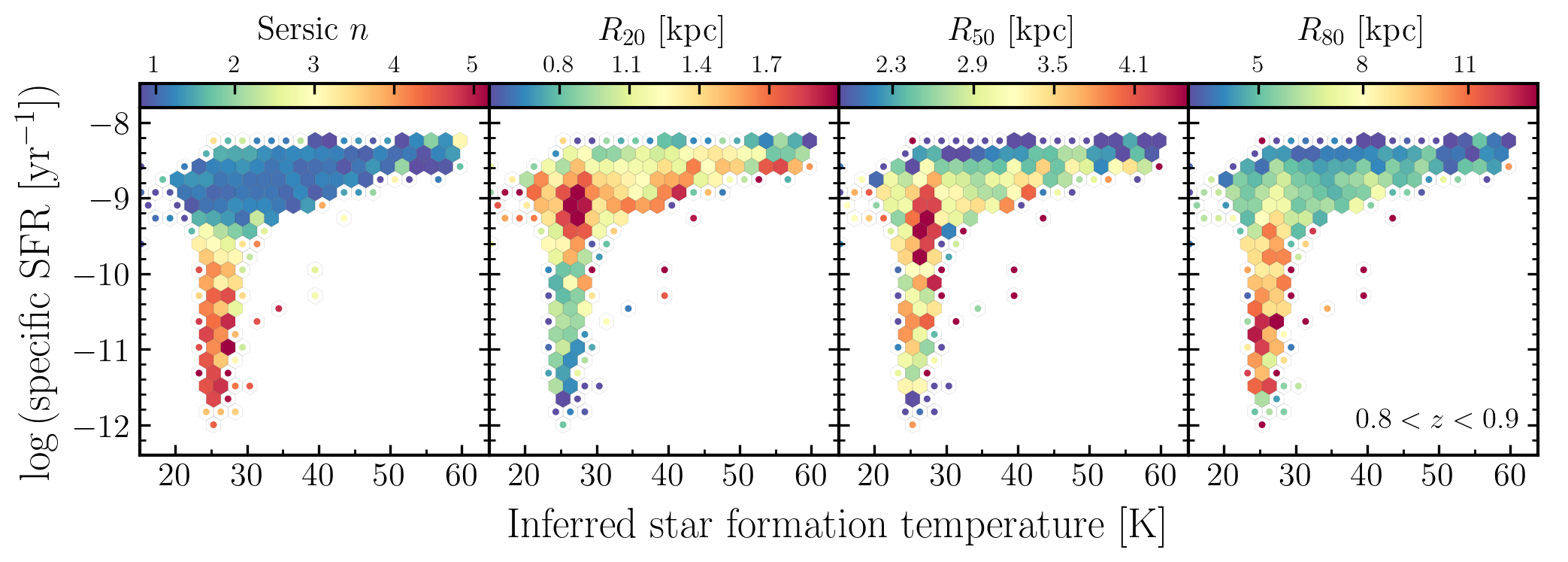}
\caption{Density plots showing the best-fit S\'{e}rsic profile properties made with {\em Hubble} WFC3 observations in the F814W filter for galaxies at $0.8<z<0.9$ (i.e. $\sim 0.44$ $\mu$m in the rest-frame).  The panels show median quantities in each $T_{\textrm{SF}}$-sSFR bin (left to right): the best-fit S\'{e}rsic index; the calculated aperture radius containing 20\% of light, $R_{20}$, using definitions from \cite{miller2019}; the best-fit effective radius, $R_{50}$; and the calculated aperture radius containing 80\% of light, $R_{80}$.  The plot indicates the following behaviors: (1) most high-sSFR systems are dominated by a disk+bulge structure or an irregular one (eg., due to mergers), while the quiescent ones have a compact luminous bulge;  (2) the change of physical sizes from $R_{20}$ to $R_{50}$ and to $R_{80}$ shows that the high-$T_{\textrm{SF}}$ objects have the smallest overall profiles, in contrast to large quiescent galaxies.  This suggests an interpretation in which the core-forming galaxies are the smallest main-sequence galaxies that grow their stellar bulge and disk over time.  Then the supposed transition to elliptical galaxies is associated with the gradual compaction of their bulge and dissipation of the disk by some instability, such as due to mergers.  Bins with fewer than 5 data points are shown as dots.}
\label{fig:sersic}
\end{center}
\end{figure*}

\begin{figure*}[!ht]
\begin{center}
\includegraphics[width=.95\linewidth]{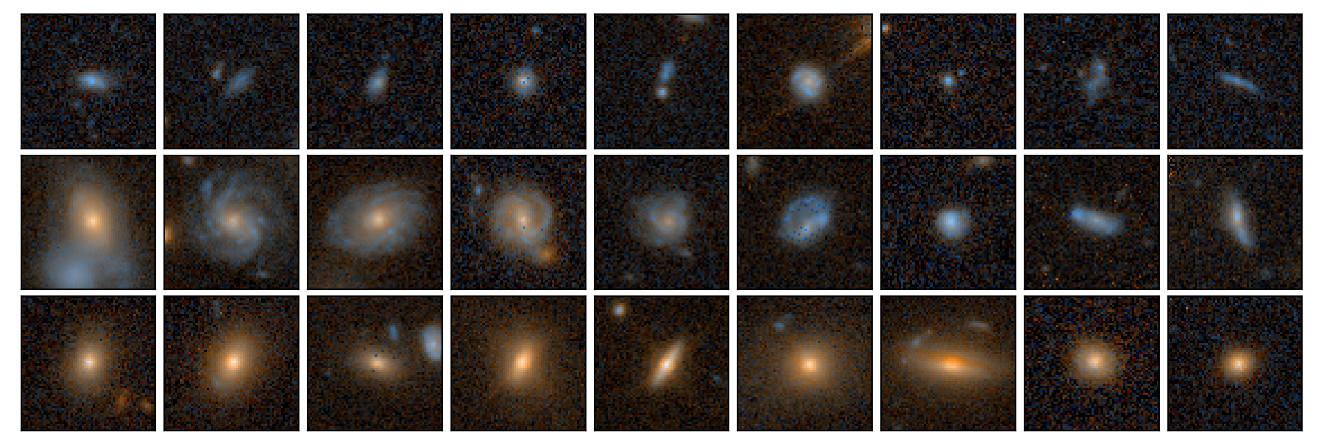}
\caption{{\em Hubble} WFC3 observations in the F814W and F160W filters for core-forming (top), typical star-forming (middle), and quiescent (bottom) galaxies randomly selected from the highest S/N objects from the COSMOS2015 survey at $z \sim 0.8$.  The $T_{\textrm{SF}}$-sSFR diagram (Figure \ref{fig:diagram}) is interpreted to indicate an evolutionary sequence from top to bottom.  The core-forming galaxies are significantly more compact and the blue color indicates a young stellar population throughout these galaxies.   Typical star-forming galaxies have blue colors and young stellar populations when unresolved, but also exhibit older populations in their centers than at larger radii.  Quiescent galaxies have old stellar populations throughout.  This morphology supports the interpretation of the $T_{\textrm{SF}}$-sSFR diagram track as an evolutionary sequence, running from core formation to typical star formation, and finally to quiescence.} 
\label{fig:cutouts}
\end{center}
\end{figure*}

This interpretation is also supported by morphological observations of COSMOS galaxies.  Much of the COSMOS field is covered by {\em Hubble} observations capable of resolving galaxies along most of the $T_{\textrm{SF}}$-sSFR diagram at $z \lesssim 1$.  Although {\em no morphological information was used} in determining best-fit parameters, producing the $T_{\textrm{SF}}$-sSFR diagram, or using that diagram to label objects as core-forming, typical star-forming, or quiescent, these categories nevertheless exhibit distinct morphology (Figs. \ref{fig:sersic}-\ref{fig:cutouts}).

The core-forming galaxies (Fig. \ref{fig:cutouts}, top) are significantly more compact than other star-forming galaxies at a redshift $z \sim 0.8$, and the blue color indicates a young stellar population throughout.  Previous studies have found that massive galaxies are typically compact at much higher redshift \citep{Bouwens2007,Bezanson2009,delaRosa2016,tacchella2016a,SchnorrMuller2021}.  It is therefore natural to interpret these $z = 0.8$ examples as much lower-redshift analogues.  Often, these are seen as a unique, high-redshift state created by extreme conditions in the first, most overdense regions to collapse.  The picture inferred from the $T_{\textrm{SF}}$-sSFR diagram is that these are instead a phase that most galaxies go through at the start of star formation.  An additional strong prediction for future dynamical studies is that the baryon distribution in these galaxies should extend well beyond the starlight into regions which are not dense enough to form stars but will do so as soon as their gas cools.

Typical star-forming galaxies (Fig. \ref{fig:cutouts}, middle) are more extended than core-forming galaxies, have blue colors and young stellar populations when unresolved, but when resolved also exhibit older populations in their centers than at larger radii.  It is already well-established that both our own Galaxy \citep{Trippe2008} and others have older stellar populations in their cores than at higher radii.  This would be a natural consequence of the model presented here.  Finally, quiescent galaxies have older stellar populations throughout, as they are no longer efficiently forming stars.  

Thus, the $T_{\textrm{SF}}$-sSFR diagram indicates not only a sequence of inferred properties, but also a morphological sequence consistent with a natural toy model derived from those properties.  Thus, much as tracks on a Hertzsprung-Russell diagram can be used to find stellar evolution sequences, tracks on the $T_{\textrm{SF}}$-sSFR diagram track can be interpreted as a galactic evolutionary sequence, running from core formation to typical star formation, and finally to quiescence. 

Although the sequence of galaxies in mass, age, and remarkably, morphology seen in Figure 1 strongly indicates an evolutionary track, there are alternative explanations informed by the appearance of similar L-shaped diagrams in both observations \citep{Barro2017} and simulations \citep{tacchella2016a,lapiner2023}. In this interpretation, galaxies simply grow along the star-forming main sequence, gradually growing in stellar surface density and effective radius following the reasonably tight relations in these quantities with stellar mass. The high-$T_{SF}$ galaxies identified here would then be explained not by a distinct stage in galactic evolution, but rather as a selection of galaxies that have recently had a burst of star formation on the timescale to which the photometry is sensitive. Both simulations \citep{Iyer2020} and observations \citep{Guo2016} suggest low-mass galaxies are preferentially bursty, which may also explain the lack of observed dark matter cusps in lower-mass galaxies \citep{Pontzen2012}. This scenario has a distinct set of predictions from the evolutionary track case, including the presence of similarly compact and blue galaxies on the star-forming main sequence with $T_{SF} \sim 25$ K with little difference in their gas structure to the high $T_{SF}$ branch. In both scenarios, the core-forming population identified here would lie in a corner of parameter space, as it lies at high $\Delta\textrm{SFR}_\textrm{MS}$ and low $\Delta\Sigma_1^Q$ on the \citet{Barro2017} diagram.

Near future observations will be able to test falsifiable predictions of these scenarios both through dynamical studies and through high-redshift studies which in the model presented in this work must consist exclusively of compact, blue, core-forming galaxies.  If those tests are consistent with predictions, it would mean that the earliest stage of star formation is indeed distinct, and that even low-redshift examples can be used to study the transition between initial assembly and subsequent evolution and probe the origins of the star-forming main sequence. \newline

The analysis in this paper is based on the COSMOS2015 catalog, available at \url{https://ftp.iap.fr/pub/from_users/hjmcc/COSMOS2015/}.  The template fits are produced by EAZY, available at \url{https://github.com/gbrammer/eazy-photoz} with templates made using FSPS, available at \url{https://github.com/cconroy20/fsps}. \newline  

The authors would like to thank Georgios Magdis, Jackson Mann, Sune Toft, and Darach Watson for helpful comments and Benjamin Magnelli for kindly providing dust temperature measurements.  The Cosmic Dawn Center (DAWN) is funded by the Danish National Research Foundation under grant No. 140.  

\bibliographystyle{aasjournal}
\bibliography{refs.bib} 



\label{lastpage}
\end{document}